\documentstyle[12pt]{article}
\begin{document}
\pagestyle{empty}
\begin{flushright}\ \\ \vskip -0.75 true in
McGill/93-9\\
TPI-MINN-93/18-T\\
May 1993\\ 
\end{flushright}
\vspace*{5mm}
\begin{center}
{\large \bf Dilepton Production in Nucleon-Nucleon Reactions
With and Without Hadronic Inelasticities}\\[10mm]
Kevin Haglin$^{a,d}$ and Charles Gale$^{b,c}$ \\[3mm] 
{\em Theoretical Physics Institute, University of Minnesota}\\
{\em Minneapolis, MN \ 55455}\\[3mm]
{\rm and}\\[3mm]
{\em Physics Department, McGill 
University, 3600 University Street}\\
{\em Montr\'eal, QC, Canada H3A 2T8} \\[3mm]
{\bf Abstract} \\
\end{center}

We calculate elementary proton-proton and neutron-proton 
bremsstrah-
lung and their contribution to the $e^+e^-$ invariant 
mass distribution.  At 4.9 GeV, the proton-proton contribution is
larger than neutron-proton, but it is small compared to recent data.
We then make a first calculation of bremsstrahlung in
nucleon-nucleon reactions with multi-hadron final states. 
Again at 4.9 GeV,
the many-body bremsstrahlung is larger than simple nucleon-nucleon
bremsstrahlung by more than an order of magnitude in the low-mass region.
When the bremsstrahlung contributions are summed with
Dalitz decay of the $\eta$, radiative decay of the $\Delta$ and from two-pion
annihilation, the result matches recent high statistics proton-proton 
data from the Dilepton Spectrometer collaboration.\\
\vspace*{10mm}

\thispagestyle{empty}\mbox{}
\vfill
\footnoterule
 {\noindent
 $^{a}$internet: haglin@hep.physics.mcgill.ca. \\
 $^{b}$internet: gale@hep.physics.mcgill.ca \\
 $^{c}$Permanent address: Physics Department, McGill University, 3600 
University Street, Montr\'eal QC, Canada H3A 2T8 \\
 $^{d}$Present address: Physics Department, McGill University, 3600 
University Street, Montr\'eal QC, Canada H3A 2T8; After
1 September 1993: Michigan 
State University, East Lansing, Michigan 48824}

\newpage\setcounter{page}{1} \pagestyle{plain}

\section{Introduction} \label{intro}
Electromagnetic probes are very well suited for studies of
strongly interacting particles under conditions ranging from
free space to nuclear matter at high energies and densities.
They carry away virtually unaltered information about the reaction
since they do not suffer strong-interaction rescattering. 
Our understanding of their properties continues to 
improve with the aid of both theoretical and experimental efforts.  For 
instance, recent high statistics experiments have been carried out at Lawrence
Berkeley Laboratory by the Dilepton Spectrometer (DLS) collaboration in
an effort to more firmly establish the mechanism of electron-positron
production in 1--5 GeV nucleon-nucleon collisions \cite{hua92}.
The result of bombarding liquid hydrogen and deuterium targets with
4.9 GeV protons yielded a $pd/pp$ ratio for $e^{+}e^{-}$ production
that was nearly mass independent and $\approx 2$.
This suggests that either bremsstrahlung is not responsible for the pair 
production yields or, if it is, that $pp$ is roughly the same as $np$.
However, conventional wisdom said that the dipole of the $np$ system
should be much more significant than the quadrupole of the $pp$ system.
Many dilepton production calculations 
have been done on $pn$, $pA$ and $AA$ systems at these energies
using as a starting
assumption that $np$ bremsstrahlung dominates over 
$pp$ \cite{cgjk87}--\cite{khag89}. 
Even a {\em complete} Feynman diagram calculation supports this 
idea \cite{khag91}.
By improving nucleon-nucleon bremsstrahlung calculations we settle
this question.  
We also generalize to some of the inelastic channels: 
those of one-, two-, and three-pion final states.  The cross section to 
produce pions in nucleon-nucleon scattering at these energies is comparable
to or even larger than elastic scattering.  One would 
therefore like to know if a proper treatment of the many-body 
electrodynamics gives a significant contribution.
The pions produced would not only carry away some fraction 
of the momentum (thereby increasing proton accelerations and therefore
radiation) but they would travel very fast since they are relatively massless
and if charged, would themselves radiate significantly.
These two complementary features could result in rather large contributions 
to the low-mass dilepton spectrum.  

Other possible mechanisms for pair production 
in these nucleon-nucleon collisions are hadronic decays,
two-pion annihilation and perhaps pion-pion bremsstrahlung. 
Pion-nucleon bremsstrahlung would
also contribute but presumably at a lower level.
Dalitz decay of $\pi^{0}\,$s contribute in a region where
experimental acceptance is low, so in this study 
we neglect them.  Contributions from
Dalitz decay of the $\eta$ and radiative decays of the low-lying nucleon
resonances might be important depending upon their 
production cross sections.  We assume two-step processes for the resonances,
namely, $NN\rightarrow NR \rightarrow 
NN\gamma^{*}$ (where $R$ might be a $\Delta$ or an $N^{*}$).
Guided by experimental data for the cross
sections and analytically continuing the expressions for real
photon production to expressions for pairs of soft 
leptons using relativistic kinematics, 
we estimate these contributions.  Two-pion annihilation was studied by 
Kapusta and Lichard in an attempt to explain possible structure near
$2m_{\pi}$ in earlier DLS data \cite{jkpl89}.  What  
they found was that when acceptance filtered, the results were not 
peaked near threshold but rather near the $\rho$ mass.
The contribution near the peak was consistent with measured dilepton
spectra when $p_{T}$ integrated.
Finally, since multiple pion production is likely at 
these energies, we estimate the contribution from $\pi\pi$
bremsstrahlung in 4.9 GeV $pp$ scattering.

The paper is organized in the following way.  In 
Sec.~\ref{softphoton} we present the soft-photon approximation with
which we calculate and
the particular continuation to finite invariant masses we use.  This will
include partial corrections to phase space which are absolutely necessary
in soft-photon approximations.  Then in Sec.~\ref{NNelastic} we apply the
formalism to nucleon-nucleon collisions with very accurate parametrizations
for the angular descriptions of elastic scattering.  
Section~\ref{NNinel} contains an application to one-, two-, and
three-pion final state many-body bremsstrahlung.  The complicated matrix
elements are approximated by their resonance structure since these
inelastic channels proceed primarily 
through formation and decay of $\Delta$-isobars.
Dalitz decay of the $\eta\,$s and radiative decays of the low-lying
nucleon resonances are estimated in Sec.~\ref{Dalitz}.  Inclusion of
a model calculation of $\pi^{+}\pi^{-}$ annihilation and its contribution
to the $e^+e^-$ mass spectrum is shown along with our $\pi\pi$ 
bremsstrahlung estimates.  Also in Sec.~\ref{Dalitz}, we show
the sum of the bremsstrahlung, $\eta$ Dalitz and $\Delta$ radiative decays 
and two-pion annihilation, as compared with dilepton yields in
proton-proton collisions from the DLS.
Then in Sec.~\ref{prediction}, we predict data for true cross sections
instead of yields and present theoretical cross sections without any
experimental (acceptance) limitations.
Finally, in Sec.~\ref{summary} we summarize our findings
concluding that many-body bremsstrahlung is the largest 
contributor to low-mass dilepton production in nucleon-nucleon 
collisions at these energies.

\section{Soft-Photon Approximation} \label{softphoton}

One very pleasant feature of calculating within a soft-photon approximation
is that the complicated algebraic structure of the diagrams separates
into two somewhat less complicated pieces.  Even if the final state
is comprised of $n >$ 2 hadrons, the matrix 
element for soft-photon production simplifies to the purely hadronic
counterpart times a multiplicative function describing the complicated
electrodynamics of the reaction \cite{leonh}.
Then as long as care is taken to approximately account for 
neglecting the momentum of the photon
in the energy-conserving delta function of phase space,
the differential cross section 
for quasi-elastic scattering while at the same time producing a virtual
photon (electron-positron pair) of invariant mass $M$ and energy $q_{0}$ is
\begin{eqnarray}
{d\sigma^{e^{+}e^{-}}\over dM} &=& 
{\alpha^{2}\over 4\pi^{4}} {1\over M q_{0}^{2}}
\widehat\sigma(s) \int \delta(q^{2}-M^{2}) \delta^{4}\left(q-(p_{+}+p_{-})
\right)d^{4}q \nonumber\\
&\ &\times{d^{3}p_{+}d^{3}p_{-} \over E_{+}E_{-}}
{R_{n}(s^{\prime}) \over R_{n}(s)},
\label{dsdm}
\end{eqnarray}
where $R_{n}$ is $n$-body phase space \cite{ebkk}, 
$s^{\prime}=s+M^{2}-2\sqrt{s}q_{0}$
and
\begin{equation}
\widehat\sigma(s) \equiv
\int \prod_{i=1}^{n} d^{3}p_{i}  \left\lbrack
{d^{3n}\sigma \over \prod_{i=1}^{n} d^{3}p_{i}} \right\rbrack
\left( q_{0}^{2} \left| \epsilon \cdot J \right|^{2}\right).
\label{shat}
\end{equation}
The square-bracketed expression in Eq.~(\ref{shat}) is the 
elastic ($n=2$) or inelastic ($n > 2$) hadronic cross section and
$q_{0}^{2}|\epsilon \cdot J|^{2}$ is the dimensionless electromagnetic
{\em weighting} representing a coherent sum of all radiation fields 
involved.  For this general process $a+b\rightarrow 1+2+\ldots +n$,
depicted in Fig.~\ref{figure1}, the electromagnetic four-current is
\begin{equation}
J^{\mu} = -{Q_{a}p_{a}^{\mu} \over p_{a}\cdot q}
-{Q_{b}p_{b}^{\mu} \over p_{b}\cdot q}+
\sum\limits_{j=1}^{n} {Q_{j}p_{j}^{\mu} \over p_{j}\cdot q},
\end{equation}
where the $Q\,$s are charges in units of the proton charge.
In this soft-photon method of calculating it is reasonable to take
an angular average over the unobserved internal photon's solid angle.  
The result of the electromagnetic {\em weighting} is \cite{khcgve93}
\vfill\eject
\begin{eqnarray}
q_{0}^{2}\left| \epsilon \cdot J\right|^{2} &=&
-(Q_{a}^{2}+Q_{b}^{2}+\sum\limits_{i=1}^{n}Q_{i}^{2}) \nonumber\\
&\ & - 2Q_{a}Q_{b}(1-\vec\beta_{a}\cdot \vec\beta_{b})
{\cal I}(\vec \beta_{a},\vec\beta_{b}) \nonumber\\
&\ &+ \sum\limits_{i=1}^{n} 2Q_{a}Q_{i}(1-\vec\beta_{a}\cdot \vec\beta_{i})
{\cal I}(\vec \beta_{a},\vec\beta_{i})
\nonumber\\
&\ &+ \sum\limits_{i=1}^{n} 2Q_{b}Q_{i}(1-\vec\beta_{b}\cdot \vec\beta_{i})
{\cal I}(\vec \beta_{b},\vec\beta_{i}) \nonumber\\
&\ &- \sum\limits_{i < j}^{n}2Q_{i}Q_{j}(1-\vec\beta_{i}\cdot \vec\beta_{j})
{\cal I}(\vec \beta_{i},\vec\beta_{j}),
\label{edotj}
\end{eqnarray}
where the function
\begin{eqnarray}
{\cal I}(\vec u, \vec v\,) &=& 
{1\over 2\sqrt{{\cal R}}}
\ln \left|{[\vec u\cdot\vec v-
u^{2}-\sqrt{{\cal R}}\ ][\vec u\cdot\vec v-
v^{2}-\sqrt{{\cal R}}\ ]\over
[\vec u\cdot\vec v- u^{2}+
\sqrt{{\cal R}}\ ][\vec u\cdot\vec v-v^{2}+
\sqrt{{\cal R}}\ ]
}\right|,
\end{eqnarray}
and the scalar
\begin{equation}
{\cal R} = (1-\vec u\cdot\vec v\,)^{2}-(1-u^{2})(1-v^{2}).
\label{duv}
\end{equation}
The $\vec\beta\,$s are particle velocities $\vec p/E$.  
This function presented in Eqs.~(\ref{edotj})--(\ref{duv})
is ultimately responsible for determining the relative strengths of
$pp$ and $np$ bremsstrahlung.  For $2 \rightarrow 2$ hadron reactions,
this function can be well approximated by a much simpler and
much used expression
when the momentum transfer is small,  e.g. it is proportional to $t$ in
$np$ scattering for $t < 4m_{N}^{2}$ \cite{cgjk87}.
But we do not make such an assumption because it does not hold generally,
a point that we believe is not fully appreciated.
We show in Fig.~\ref{figure2} this electromagnetic
{\em weighting} for 4.9 GeV  $np$ and $pp$ reactions 
as a function of center-of-mass scattering angle.
As we see, $pp$ is suppressed only at the poles and 
rises to some rather large value for intermediate angles. For 
scattering angles in the forward hemisphere
$pp$ is larger than $np$.  This is not inconsistent with the assumption
that has prevailed in discussions of bremsstrahlung about $np$ dominating
$pp$ since the assumption was based on a non-relativistic argument.
At 100 MeV kinetic energy, for 
instance, the maximum value of the $pp$ weighting (at $\cos \theta = 0$)
is $\approx 4.0\times 10^{-3}$, whereas the $np$ weighting at that angle 
is $\approx 7.0\times 10^{-2}$.  For brevity, we do not show results
at other energies, but already at 1 GeV the 
$np$ domination breaks down.
In the next section we
parametrize $pp$ and $np$ angular distributions guided by
experimental data, weight them
with the functions plotted in Fig.~\ref{figure2} by
evaluating Eq.~(\ref{dsdm}), and obtain the
$e^{+}e^{-}$ invariant mass spectra.

\section{Nucleon-Nucleon Bremsstrahlung} \label{NNelastic}

The absolute strength of nucleon-nucleon bremsstrahlung
depends, among other things, on the differential elastic cross 
section $d\sigma /dt$.  By definition, this is symmetric for $pp$ 
scattering at any energy, whereas,
it is slightly asymmetric for $np$ at energies of interest to this 
study.  This asymmetry can suppress the $np$ bremsstrahlung contribution to 
dilepton production by a factor 4 for 4.9 GeV reactions as compared 
with calculations using a symmetric $d\sigma /dt$ \cite{winck,keywest}.  
Clearly what is needed in order to minimize uncertainty in normalization
are very accurate parametrizations 
of the experimental data.  Functional forms that do quite nicely for 
energies 1--5 GeV are presented here.  First for $pp$ at 4.9 GeV we use
\begin{eqnarray}
{d\sigma_{pp} \over d\Omega} &=&
\exp\left\lbrace \Theta(\theta-\theta_{0})\Theta\left(
(\pi-\theta_{0})-\theta\right) \left(A + B|\theta-\pi/2|^{\gamma}
\right) \right. \nonumber\\
&\ & + \Theta(\theta_{0}-\theta)(C-D|\theta|^{\delta}) \nonumber\\
&\ & + \left. \Theta\left(\theta-(\pi-\theta_{0})\right)
(C-D|\pi-\theta|^{\delta}) \right\rbrace
\label{ppelcross}
\end{eqnarray}
where $\Theta$ is the Heavyside-Step function,
\begin{eqnarray}
A &=& \log {d\sigma_{pp}(\pi/2) \over d\Omega} \nonumber\\
B &=& |\theta_{0}-\pi/2|^{-\gamma}
\log \left( { {d\sigma_{pp}(\theta_{0})/d\Omega} \over 
              {d\sigma_{pp}(\pi/2)/d\Omega} } \right) \nonumber\\
C &=& \log {d\sigma_{pp}(0) \over d\Omega} \nonumber\\
D &=& |\theta_{0}|^{-\delta}
\log \left( { {d\sigma_{pp}(0)/d\Omega} \over 
              {d\sigma_{pp}(\theta_{0})/d\Omega} } \right),
\end{eqnarray}
and $\gamma = 2.3$, $\delta=1.6$, $\theta_{0}=\pi/6$,
$d\sigma_{pp}(0)/d\Omega=80$ mb/sr,
$d\sigma_{pp}(\theta_{0})/d\Omega=0.9$ mb/sr and 
$d\sigma_{pp}(\pi/2)/d\Omega=5.0\times 10^{-3}$ mb/sr;
and then for $np$ scattering again at 4.9 GeV we use
\begin{eqnarray}
{d\sigma_{np} \over d\Omega} &=&
\exp\left\lbrace \Theta(\theta_{0}-\theta) 
\left(A + B|\theta-\theta_{0}|^{\gamma}
\right) \right. \nonumber\\
&\ & + \left. \Theta\left(\theta-\theta_{0}\right)
(C+D|\theta-\theta_{0}|^{\delta}) \right\rbrace,
\label{npelcross}
\end{eqnarray}
\begin{eqnarray}
A &=& \log {d\sigma_{np}(\theta_{0}) \over d\Omega} \nonumber\\
B &=& \theta_{0}^{-\gamma}
\log \left( { {d\sigma_{np}(0)/d\Omega} \over 
              {d\sigma_{np}(\theta_{0})/d\Omega} } \right) \nonumber\\
C &=& \log {d\sigma_{np}(\theta_{0}) \over d\Omega} \nonumber\\
D &=& |\pi-\theta_{0}|^{-\delta}
\log \left( { {d\sigma_{np}(\pi)/d\Omega} \over 
              {d\sigma_{np}(\theta_{0})/d\Omega} } \right),
\end{eqnarray}
and $\gamma = 2.3$, $\delta=2.5$, $\theta_{0}=99$ degrees,
$d\sigma_{np}(0)/d\Omega=6.27\times 10^{2}$ mb/sr,
$d\sigma_{np}(\theta_{0})/d\Omega=1.62\times 10^{-5}$ mb/sr and 
$d\sigma_{np}(\pi)/d\Omega=16.2$ mb/sr.

Comparisons are made with $np$ \cite{{jls77},{elm71},{mlp70}}
and $pp$ \cite{{iam74},{rck71}}
data in Figs.~\ref{figure3}a and \ref{figure3}b.
Upon integrating these expressions over the scattering
angle, or equivalently over momentum transfers $-(s-4m_{N}^{2})
\le t \le 0$, we obtain cross sections of 11.9 mb and 11.1 mb 
for $pp$ and $np$, respectively.  Recall, there is a factor $1/2$ in
the $pp$ case to avoid double counting of the identical particles.

In order that we might compare to published data on dilepton
production, we include the DLS acceptance filter to calculate
a yield.  The results are then acceptance filtered but not acceptance 
corrected~\cite{hua92}, and they have units of cross section.
When Eqs.~(\ref{ppelcross}) and (\ref{npelcross}) are used in
Eq.~(\ref{shat}), and in turn in Eq.~(\ref{dsdm}) (with acceptance
filter and weight included), the resulting
spectra differ by a constant factor of $\approx 1.5$.  That the
spectra differ by a constant is a consequence of our soft photon
approximation.    
We show the yields in Fig.~\ref{figure4}, where the comparison is
absolute.  Proton-proton 
bremsstrahlung is stronger than neutron-proton at this energy!  
If we construct the ratio of 
$\sigma(np+pp)/\sigma(pp)$, we find it to be 1.67.  Therefore, 
it is not the measured ratio of order 2 that eliminates bremsstrahlung
from being responsible for the $e^{+}e^{-}$ yields;
one must compare the distributions separately. 
Upon doing so, one if forced to conclude that simple bremsstrahlung
is not responsible for the observed $pp$ yields.
One naturally thinks to estimate the inelastic
channels' contributions to $e^{+}e^{-}$ production.  After all,
at 5 GeV the cross section is mostly inelastic.

\section{Multiple-Hadron Final States} \label{NNinel}

The cross sections to produce
up to and including five pions have been measured in 5.5 GeV/c
proton-proton collisions \cite{alex}.   The one, two and perhaps three
pion final states are important for our considerations here.  Their
cross sections are 10.8, 11.4, and 11.7 mb, respectively.  At the same
time the elastic channel's cross section is 11.3 mb.  So clearly,
these inelastic channels cannot be ignored.  
What is needed are the $NN\rightarrow NN\pi$, 
$NN\rightarrow NN\pi\pi$, and
$NN\rightarrow NN\pi\pi\pi$ matrix elements.  One could of course perform
a field theory calculation to get a fairly complete description of them.
The interactions of $\pi$ and $\rho$ mesons coupling to a nucleon and 
a delta are well known.  However, the $\Delta$ propagator 
presents some technical difficulty making
a full-blown calculation extremely lengthy.  Furthermore, even a detailed
one-boson-exchange calculation is not guaranteed to produce correct 
angular behavior for the differential cross section.
We pursue a different approach.
The process $pp\rightarrow np\pi^{+}$ proceeds primarily through 
$\Delta^{+}$ or $\Delta^{++}$ excitation of the nucleon,
with subsequent decay into a nucleon and the positively charged pion.
The corresponding diagrams are shown in Figs.~\ref{figure5}a and 
\ref{figure5}b, with
exchange diagrams not explicitly drawn.  To this order, there are also
effects from $\Delta^{0}$ excitation as in Fig.~\ref{figure5}c.
Even though the matrix element for these diagrams contains influences from 
force-mediating meson-exchange, the resonance structure of the 
$\Delta$ propagator alone accounts reasonably well
for what is observed in terms of invariant mass distributions.
Therefore, we write the matrix element 
for the diagrams in Fig.~\ref{figure5} as
\begin{eqnarray}
{\cal M}_{pp\rightarrow np\pi^{+}} &\propto&
{1\over \sqrt{2}} { (p_{1}+p_{3})^2-(m_{N}+m_{\pi})^2 \over 
  (p_{1}+p_{3})^2-m_{\Delta}^{2}+im_{\Delta}\Gamma_{\Delta} } \nonumber\\
&-&
{7\over 3\sqrt{2}}
{ (p_{2}+p_{3})^2-(m_{N}+m_{\pi})^2 \over 
  (p_{2}+p_{3})^2-m_{\Delta}^{2}+im_{\Delta}\Gamma_{\Delta} },
\end{eqnarray}
where $a+b\rightarrow 1+2+3$ corresponds to particles $p+p\rightarrow
n+p+\pi^{+}$.  With this matrix element, we are neglecting the neutral
excitation diagram.  We have verified
that its inclusion is not significant to the final dilepton spectrum,
at least within this approximation.  Keeping complexity to an absolute
minimum while capturing the essential physics, we choose to
neglect contributions from diagrams like the one in Fig.~\ref{figure5}c. 
Then the cross section can be calculated by
\begin{equation}
d\sigma = {2m_{N}^{4} \over \sqrt{s(s-4m_{N}^{2})}} {|{\cal M}|^{2}
\over (2\pi)^5}
\delta^{4}(p_{a}+p_{b}-p_{1}-p_{2}-p_{3}) 
{d^{3}p_{1} \over E_{1}}{d^{3}p_{2} \over E_{2}}{d^{3}p_{3} \over 2E_{3}}.
\end{equation}
This reduces to an integration over the four 
essential final state variables describing phase space.
The normalization of the matrix element is adjusted in order to give a 
cross section of $\sigma(pp\rightarrow np\pi^{+}) = 8.03$ mb.  
Similar approximations
are done for the matrix elements of the processes
$pp \rightarrow pp\pi^{0}$, $np \rightarrow nn\pi^{+}$,
$np \rightarrow np\pi^{0}$ and $np \rightarrow pp\pi^{-}$.
The cross sections we use are: 2.77 mb for $pp\rightarrow pp\pi^{0}$,
8.03 mb for $np\rightarrow nn\pi^{+}$ and finally, 2.77 mb for both
$np\rightarrow np\pi^{0}$ and $np\rightarrow pp\pi^{-}$.
Future model calculations are needed to improve on these matrix
elements, but for first approximation they are sufficient.  

With these hadronic one-pion-production cross sections, we insert into
Eq.~(\ref{shat}) and subsequently into Eq.~(\ref{dsdm})
in order to calculate each channel's contribution to dilepton
production.  Three-body phase space can be written as
\begin{equation}
R_{3}(s) = \int\limits_{(m_{N}+m_{\pi})^2}^{(\sqrt{s}-m_{N})^2}
ds_{2} R_{2}(s,s_{2},m_{N}^{2})R_{2}(s_{2},m_{N}^{2},m_{\pi}^{2}).
\end{equation}
When Eq.~(\ref{dsdm}) is evaluated for the $n=3$ case under 
consideration and with the acceptance filter included upon 
integration over dilepton
$p_{T}$ and rapidity, a yield is obtained
that can be directly compared with the simple $pp$ bremsstrahlung
presented earlier.  In Fig.~\ref{figure6}
we show the two single-pion final state
contributions superimposed with simple bremsstrahlung.
The largest contributor is the channel 
$pp\rightarrow np\pi^{+}e^{+}e^{-}$.  For low
invariant masses, it is larger than simple bremsstrahlung by nearly an 
order of magnitude.  This many-body contribution has been ignored
in previous calculations for dilepton production because
it was assumed small.  It is clear that such an omission is not justified. 
The same analysis applied to the $np$ channels gives similar results.
Charged nucleon reaction partners are of course important for bremsstrahlung.
But we have discovered that since the inelastic cross sections are
comparable in size to elastic, having a charged pion in the final state
significantly boosts the radiation.

We then proceed to estimate the contributions from the two-pion final 
states.  There are four channels each for 
$pp$ and $np$ scattering.  The same 
kind of approximations are made in order to arrive at the matrix elements.
This time, however, there are two $\Delta$ excitations and the isospin
factors are different.  The general structure of the diagrams is
shown in Fig.~\ref{figure7}.  Again, we are neglecting diagrams that 
contain delta excitations on the initial lines since they are not
essential.  The matrix element we use for the process in 
Fig.~\ref{figure7} is
\begin{eqnarray}
{\cal M}_{pp\rightarrow pp\pi^{+}\pi^{-}} &\propto&
 { (p_{1}+p_{3})^2-(m_{N}+m_{\pi})^2 \over 
  (p_{1}+p_{3})^2-m_{\Delta}^{2}+im_{\Delta}\Gamma_{\Delta} }
{ (p_{2}+p_{4})^2-(m_{N}+m_{\pi})^2 \over 
  (p_{2}+p_{4})^2-m_{\Delta}^{2}+im_{\Delta}\Gamma_{\Delta} } \nonumber\\
&+&
 { (p_{2}+p_{3})^2-(m_{N}+m_{\pi})^2 \over 
  (p_{2}+p_{3})^2-m_{\Delta}^{2}+im_{\Delta}\Gamma_{\Delta} } \nonumber\\
&\ & \times
{ (p_{1}+p_{4})^2-(m_{N}+m_{\pi})^2 \over 
  (p_{1}+p_{4})^2-m_{\Delta}^{2}+im_{\Delta}\Gamma_{\Delta} },
\end{eqnarray}
where $p_{1}$ and $p_{2}$ label the protons and $p_{3}$($p_{4}$)
labels the positive(negative) pion.
Similar approximations give the matrix elements for $pp\rightarrow pp\pi^{0}
\pi^{0}$, $pp\rightarrow np\pi^{+}\pi^{0}$, $pp\rightarrow nn\pi^{+}
\pi^{+}$, and $np\rightarrow np\pi^{+}\pi^{}$, $np\rightarrow np\pi^{0}
\pi^{0}$, $np\rightarrow pp\pi^{-}\pi^{0}$ and $np\rightarrow nn\pi^{+}
\pi^{0}$.  
For simplicity, we assume an isospin equivalence forcing a numerical 
value of 2.85 mb for the cross sections of all two-pion final states.
Measured $pp$ channels differ from this by only a small amount, so
it is not unreasonable.

The phase space correction this time is $R_{4}(s^{\prime})/R_{4}(s)$,
where
\begin{eqnarray}
R_{4}(s) &=& \int\limits_{s_{2}^{-}}^{s_{2}^{+}}
\int\limits_{m_{2}^{-}}^{m_{2}^{+}}
ds_{2}dm_{2} R_{2}(s,s_{2},m_{2})R_{2}(s_{2},m_{N}^{2},m_{\pi}^{2}) 
R_{2}(m_{2},m_{N}^{2},m_{\pi}^{2}), \ \ \
\end{eqnarray}
with $s_{2}^{-}=(m_{N}+m_{\pi})^2$, $s_{2}^{+}=(\sqrt{s}-m_{N}-m_{\pi})^2$,
$m_{2}^{-}=(m_{N}+m_{\pi})^2$ and $m_{2}^{+}=(\sqrt{s}-\sqrt{s_{2}})^2$.
Naturally, this correction has a tendency to suppress higher massed
pairs.  The suppression with two pions in the final state is greater 
than it is having just a single pion.  Carrying out the necessary phase 
space integration we find the resulting contributions to $e^{+}e^{-}$
yields from two-pion final states to be intermediate between simple
$pp$ bremsstrahlung and the largest single-pion channel.
There are enough channels, however, so that when added together the 
low-mass contribution is slightly larger than that from single-pion
final state bremsstrahlung.

Finally, there is the issue of $m>2$ pion final states.  Note that
at 4.9 GeV there is enough phase space to allow as many as 12 pions
in the final state.  The cross section
to produce three pions in $pp$ scattering at 5.5 GeV/c is 11.7 mb.
By doing a proper treatment of the five-body electrodynamics and by
approximating the matrix elements in our by now familiar fashion, we find the
contribution important only to the lowest two or three invariant mass
bins.  Specifically, using a cross section of 2.34 mb
for the channel $pp\rightarrow np\pi^{+}\pi^{+}\pi^{-}$, we find
$dn^{e^{+}e^{-}}/dM 
\approx 8.0\times 10^{-5} \ \mu b/$GeV for $M=0.075$ GeV.  Admittedly this 
seems relatively large; but the five-body
phase space correction very strongly suppresses the distribution for 
increasing invariant mass, 
bringing it down to $4.0 \times 10^{-6}\ \mu b/$GeV at $M=0.500$ GeV.  
There are five charge configurations for three-pion final states which 
must all be included.  Assuming charge independence for the cross sections, 
we arrive at our aforementioned value of 2.34 mb for each of these 
channels.  The relative smallness of the cross sections for four- and 
five-pion production justifies ignoring them here.  

In Figure~\ref{figure8} we show the sum of zero-,
one-, two-, and three-pion final-state hadronic bremsstrahlung.
One should compare to data only in the low-mass region where this 
soft-photon approximation is best.  Upon doing so, we may conclude that
these inelastic channels are responsible for much of the low-mass dilepton
yield.  A previous dilepton calculation which included a comparison 
between the soft-photon approximation and a Feynman-diagram method in
4.9 GeV $np$ scattering showed the soft-photon result smaller than 
the diagram calculation by a factor of 6 at invariant mass 300 MeV.
The two results converged to the same curve (as they must)
in the limit $M\rightarrow 0$ \cite{khag91}.  So one is encouraged that 
bremsstrahlung might even account for the yields at low-to-intermediate 
masses.

\section{Other sources} \label{Dalitz}

Hadronic decays are the next most likely candidate for a
mechanism of significant dilepton production.  
Of the possible
hadrons that might be excited or produced in nucleon-nucleon
collisions at energies $\sim$ 1 GeV, we find that the $\eta$ and
the $\Delta$ contribute the most.  The rho and omega are 
somewhat below the delta, so we ignore them here.
The delta has a direct decay channel $\Delta \rightarrow N\gamma$ with
measured branching ratio $0.6\%$.  The photon might not satisfy the
Einstein condition $q^{2}=0$, but instead appear as a virtual (massive)
photon.  The precise mathematical formulation of the analytical continuation
we use is identical to the calculation published first by Gale and 
Kapusta in Ref.~\cite{cgjk89}.\footnote{The acceptance filter we use 
is Version 1.6, a later release than the one used in Ref.~\cite{cgjk89}.}
The $\Delta$ production
cross section we use is the measured value of 5.0 mb.  We have also 
checked the contributions from radiative decay of the
first few low-lying $N^{*}\,$s and found them to be small compared to
the $\Delta$.
Calculation of the eta proceeds in the same way as
the estimate from Ref.~\cite{hua92} with
one exception.  We use the upper limit for the eta production cross section:
0.5 mb.  Then a differential cross section is assumed which is gaussian
in laboratory rapidity and exponential in transverse mass.  For more details
see Ref.~\cite{hua92}.


At 4.9 GeV, the cross section to produce pions is larger than elastic
scattering.  The 1, 2, 3, and 4 pion production cross sections
in $pp$ scattering are 10.8, 11.4, 11.7 and 1.6 mb, respectively.  Since
the reaction zone has finite extent, there is clearly some nonzero 
probability for pion-pion rescattering.  Furthermore, vector dominance 
would suggest that in such scattering the rho resonance accounts for
most of the cross section.  It is well known that the rho has an
electromagnetic decay channel, so it is reasonable to estimate the
contribution to dilepton production through two-pion annihilation.  This
problem has been solved by Kapusta and Lichard in a search for structure
in the dilepton mass spectrum \cite{jkpl89}.  They found a structure
appearing at the rho mass (instead of near $2m_{\pi}$)
when the distribution was acceptance corrected. 
For details of this kinetic theory calculation see Ref.~\cite{jkpl89}.

Pion-pion rescattering with bremsstrahlung will also contribute
to dilepton production.  
Although the contribution from 
$\pi\pi \rightarrow \pi\pi e^{+}e^{-}$ might be rather 
large \cite{khcgve93}, we must
fold in some probability for two (final-state) pions to scatter, and
then sum over all possible scattering energies and charge configurations.
This requires complete knowledge of the dynamics.  However,
an estimate can be obtained from the following relation
\begin{equation}
\left({d \sigma_{\pi\pi}^{e^{+}e^{-}} \over dM}\right)_{\rm rescattering} =
\int\limits_{2m_{\pi}+M}^{\sqrt{s}-2m_{N}} \hskip -0.125 true in
dm \; \;
{dP\over dm} 
\left({d \sigma_{\pi\pi}^{e^{+}e^{-}} \over dM}\right)_{\rm brems.},
\label{rescatter}
\end{equation}
where the probability for a pion to rescatter off another (pion)
with invariant mass squared $m^{2} = (p_{\pi_{1}}+p_{\pi_{2}})^{2}$ in 
a 4.9 GeV $pp$ collision is
\begin{equation}
{dP\over dm} \approx \left. \left({d\sigma_{\pi^+\pi^-}^{e^{+}e^{-}} \over  
dm} \right) \right/  
\sigma_{\pi^+\pi^-}^{e^{+}e^{-}}.
\label{prob}
\end{equation}
The annihilation cross section $\sigma_{\pi^+\pi^-}^{e^{+}e^{-}}$, including 
a vector dominance form factor, is taken from Eq.~(15) of Ref.~\cite{cgjk87}.
Finally, the
kinetic theory calculation of Kapusta and Lichard gives an estimate
of the differential annihilation cross section \cite{jkpl89},
$d\sigma_{\pi^+\pi^-}^{e^+e^-}/dm$. Having the probability for 
rescattering, we take the pion-pion 
bremsstrahlung expression developed within a soft photon approximation
in Ref.~\cite{khcgve93} and evaluate the (invariant energy)
integral in Eq.~(\ref{rescatter}).  In practice, this is a finite sum
$\sum_{i} \Delta m_{i}(dP/dm) \times (d\sigma/dM)_{\rm brems}$.
The resulting contribution to dilepton production 
is comparable to simple proton-proton bremsstrahlung in the low
mass region and intermediate masses while it drops rapidly 
for pair masses above $\approx$ 600 MeV.
Note that this represents
a lower bound on the pion-pion bremsstrahlung contribution
for two reasons.  Firstly, we have
neglected other charge configurations and secondly, the probability 
calculated with Eq.~(\ref{prob}) has experimental acceptance contaminating
the numerator.  The contamination reduces the probability in the mass region
near threshold and should, in principle, not be there.  However, the
effect is very small.  Since the pion-pion result is comparable to simple
proton-proton bremsstrahlung at this energy, 
it is clearly not dominant.  On the other hand, it is not insignificant 
either and should be considered in future calculations.  Without performing
a detailed numerical simulation, this is as far as we will take such
estimates.

The dilepton spectra arising from (total) bremsstrahlung, hadronic
decays and from two-pion annihilation are all presented in Fig.~\ref{figure9}.
Dalitz decay of the eta peaks at a mass $\approx$ 325 MeV 
and at a level slightly above bremsstrahlung.  Therefore,
it is a crucial ingredient in the final spectrum.
The delta result is bimodal since, like all our results, it is
acceptance filtered.  The two peaks appear at 200 MeV
and at the rho mass.  From this it seems the delta is less important than
the eta since the peak at 200 MeV is well below bremsstrahlung, and the
other peak is masked behind the contribution from two-pion annihilation. 

\section{Cross Sections} \label{prediction}

In the previous sections we have presented dilepton
yields since cross sections have not yet been published.
Now we include a short section 
in which we predict the cross section data for 4.9 GeV proton-proton
collisions, since they are forthcoming.  The dominant processes,
mechanisms and calculational methods are exactly the same as we have
discussed in the previous sections, with one exception.  Here 
we perform an acceptance correction, and therefore the answer is 
a true cross section~\cite{explainw}.  In Fig.~\ref{figure10} 
we show the resulting invariant mass distributions of true cross sections
for bremsstrahlung, hadronic decays and two-pion annihilation.  The prediction
is absolute.  

Another useful comparison is to present cross sections for the various 
mechanisms without any experimental limitations included, i.e. without any
acceptance effects.  This {\em theoretical} comparison no longer
depends on a particular experimental apparatus.  
In Fig.~\ref{figure11} we show such
a comparison.  The hadronically inelastic channels contribute significantly
in the low-mass region.

\section{Summary} \label{summary}

We have shown by properly calculating the electromagnetic interference 
in $np$ and $pp$ scattering and by accurately parametrizing
the differential cross sections, that the $e^{+}e^{-}$ invariant mass
distribution from $pp$ bremsstrahlung is larger than $np$ at 4.9 GeV.
The single, double, and triple pion production hadronic cross sections were 
calculated while approximating the matrix elements by their gross 
properties of $\Delta$ formation and decay.
Knowing the cross sections for these inelastic channels in 
$pp$ scattering, we calculated
their contribution to the dilepton mass spectrum.  The single-pion 
production channel (with a charged pion) contributed the most to 
low-mass pairs.  
Dalitz decay of the eta contributed roughly at the
same level in the narrow window near its peak.
Finally, the sum of contributions from simple
and inelastic hadronic bremsstrahlung, Dalitz decay of the
$\eta$, radiative decay of $\Delta$, and from 
$\pi^{+}\pi^{-}$ annihilation, satisfactorily describes the 
measured $pp$ distribution.  We conclude that bremsstrahlung is 
indeed the largest source of low-mass dileptons in these 4.9 GeV $pp$
collisions, but that it comes from many-body bremsstrahlung.

Although the agreement with experimental data is quite good, it
is not perfect.  We do not attempt in this paper to achieve a closer fit.
Our main goal was to establish the importance of the channels with 
multi-particle final states in electromagnetic radiation calculations.
We have neglected several factors that can be important for
precise quantitative interpretation of the experimental
data.  We have no effects of form factors in our dilepton emission
many-body cross sections.  In this sense, the curves shown in this work
represent a lower bound only.  However, we have also neglected 
interference effects in the radiative resonance decays.  These should 
be carefully examined since they have been shown to be of some 
importance in both dilepton \cite{khag91} and photon \cite{song92} 
calculations.   Radiative decays from multiple $\Delta$ excitations 
have also been neglected here.  Potentially important effects in the
pion-nucleon channels have not been treated either.  Calculation
of these rescattering effects requires complete knowledge of the kinematics
which is outside the scope of this paper.

The hadronic inelastic channels will now play a major role in the 
interpretation and understanding of experimental data at these energies.
An immediate consequence of our study is that high energy heavy-ion
data will require the use of state-of-the-art many-body numerical
simulations, where multi-particle final-state channels 
and their appropriate electromagnetic weighting must be included.
It is of great importance to study the rise of the contributions with 
the incident kinetic energy to carefully map out the threshold effects.  
We eagerly await the upcoming data from the DLS and from HADES, the 
European dilepton collaboration. While it is true that the proliferation 
of new channels will complicate the many-body problem, we can on the 
other hand state that it will no doubt contribute to its richness.

\section*{ Acknowledgements }

We both wish to acknowledge the hospitality of the
Theoretical Physics Institute and Physics 
Department at the University of Minnesota where the initial stage of this
work was done.  Also, we acknowledge useful discussions with Guy Roche.
Our research is supported in part by the Natural 
Sciences and Engineering Research Council of Canada, a NATO 
collaborative research grant and the FCAR fund of the Qu\'ebec government.
For this work we also acknowledge the U.S. Department of Energy, grant
number DOE/DE-FG02-87ER40328.





\section*{ Figure captions }

\begin{enumerate}

\item Hadron-hadron collision with an $n$-particle final state where the
arrows indicate momentum flow.  A coherent sum of radiation for all charged 
external lines is computed.
\label{figure1}

\item Dimensionless electromagnetic {\em weighting} as it depends
on the center-of-mass scattering angle in 4.9 Gev 
nucleon-nucleon collisions. 
\label{figure2}

\item Parametrizations of $np$ and $pp$ differential cross sections
in (a) and (b), respectively.  Experimental data are from 
(a) Refs.~\cite{{iam74},{rck71}}, and
(b) Refs.~\cite{{jls77},{elm71},{mlp70}}.
\label{figure3}

\item Dilepton yield from simple $pp$ and $np$
bremsstrahlung as compared with DLS data from Ref.~\cite{hua92} for
4.9 GeV proton-proton (open circles) and proton-deuteron (solid squares)
inclusive reactions.  
Our $pp$ result is shown as the dashed histogram, $np$ is shown as the dotted
histogram, and an approximate $pd = pp + np$ is shown as the
solid histogram.
\label{figure4}

\item Pion production diagrams proceed through $\Delta^{+}$ formation
and decay as in (a), through $\Delta^{++}$ formation
and decay as in (b) and through $\Delta^{0}$ excitation as in (c).
The exchanged meson $\phi$ might be a pion,
a rho-meson, or some other boson.
\label{figure5}

\item Dilepton yields obtained from $NN \rightarrow NN\pi$ 
bremsstrahlung as compared with simple nucleon-nucleon bremsstrahlung, and 
with $pp$ data.
\label{figure6}

\item Two-pion-production diagrams proceed via $NN\rightarrow\Delta\Delta
\rightarrow NN\pi\pi$.
\label{figure7}

\item The sum of bremsstrahlung contributions from $NN$, $NN\pi$,
$NN\pi\pi$, and $NN\pi\pi\pi$ final states in $pp$ scattering.  Data are 
again $pp$ results from the Bevalac.
\label{figure8}

\item Dilepton yields from various mechanisms:
bremsstrahlung is presented as solid squares, radiative decay 
of the $\Delta$
is presented as solid triangles, Dalitz decay of the $\eta$ is shown as open
squares, two-pion annihilation is shown as open diamonds,
and finally, the
sum of all contributions shown as the solid histogram. 
\label{figure9}

\item Cross section (invariant mass distributions) from various mechanisms:
bremsstrahlung is presented as solid squares, 
radiative decay of the $\Delta$
is presented as solid triangles, Dalitz decay of the $\eta$ is shown as open
squares, two-pion annihilation is shown as open diamonds
and finally, the sum of all contributions shown as the solid histogram.
\label{figure10}

\item Cross sections without any experimental acceptance included.
Labels indicate the following mechanisms: (a) simple proton-proton 
bremsstrahlung in the solid line,
(b)--(e) are $pp\pi^{0}$ in the dotted line, $np\pi^{+}$ in the
short-dashed line, $pp\pi^{+}\pi^{-}$ in the dot-dashed line
and $np\pi^{+}\pi{+}\pi^{-}$ final-state bremsstrahlung in the long-dashed
line; (f) is eta Dalitz decay and finally,
radiative Delta decay is shown in curve (g).
\label{figure11}

\end{enumerate}

\vfill \eject


\begin{thebibliography}{99}
\bibitem{explainw} The DLS acceptance filter determines detectable phase
space for the lepton pair as well as a weighting $1/w$ for each region
in this space.  Yields are calculated with the weighting multiplying 
$d\sigma/dyd^{2}p_{T}$ upon integration over dilepton phase space, 
while cross sections are calculated without.
\end{thebibliography}
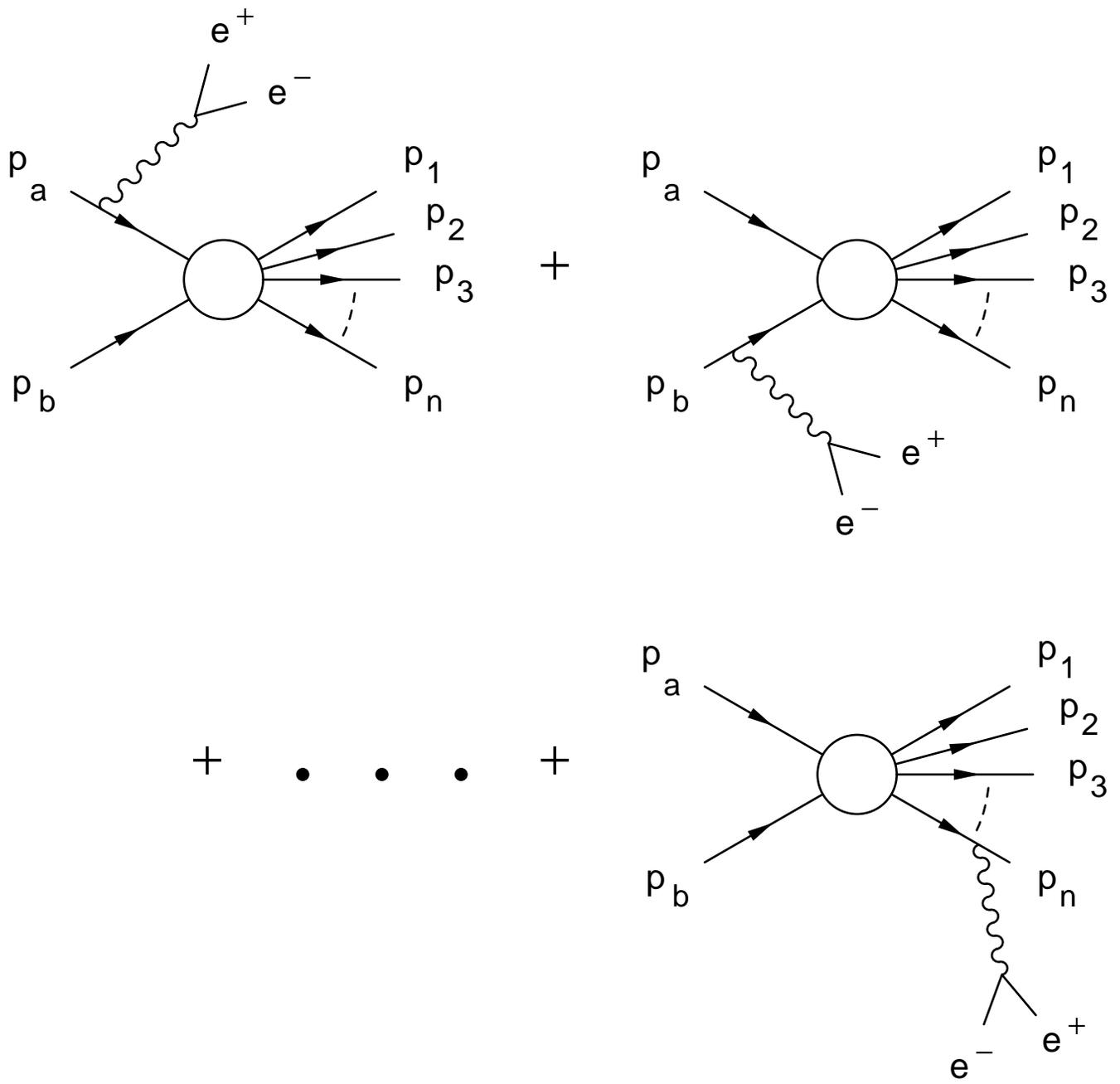
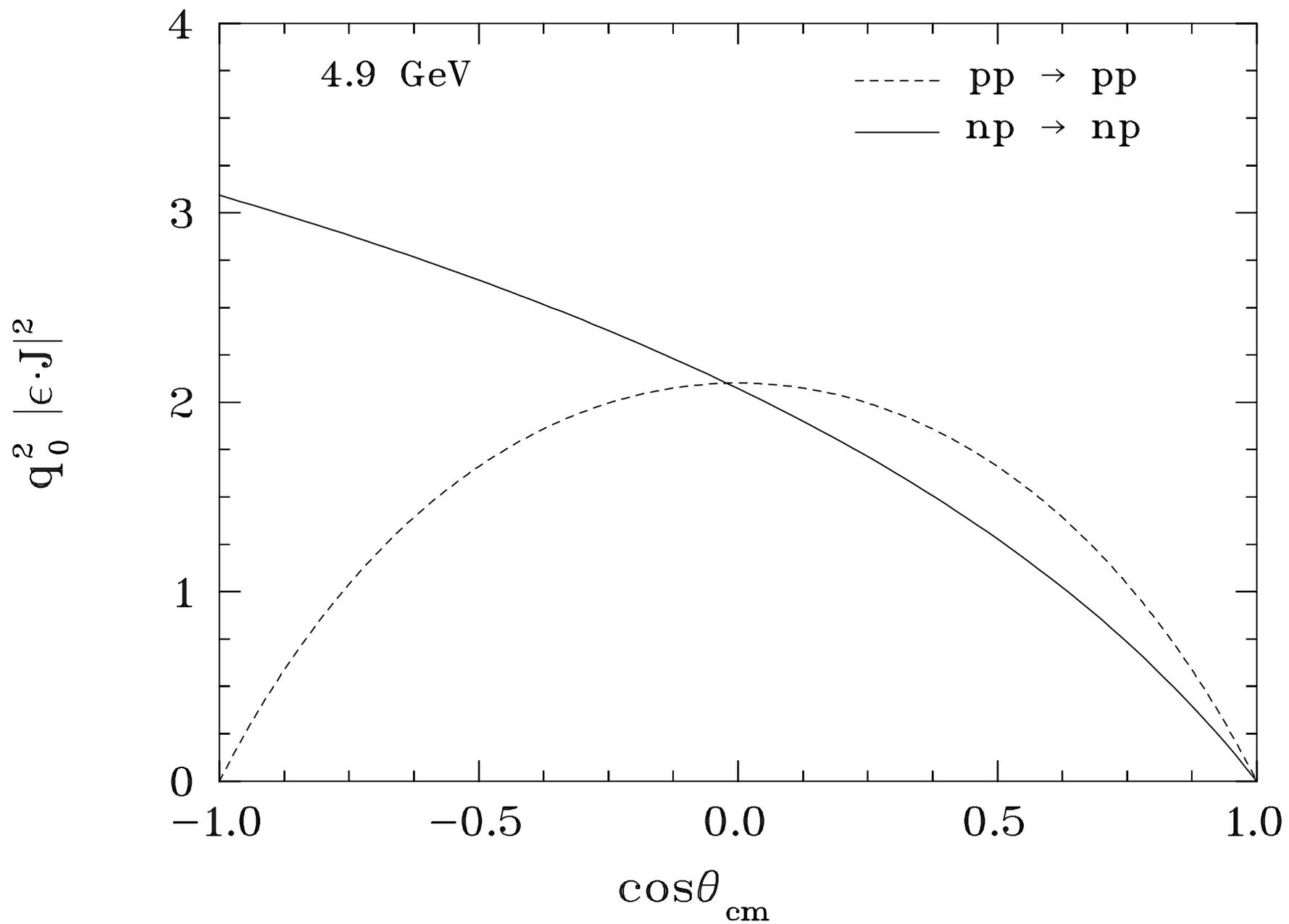
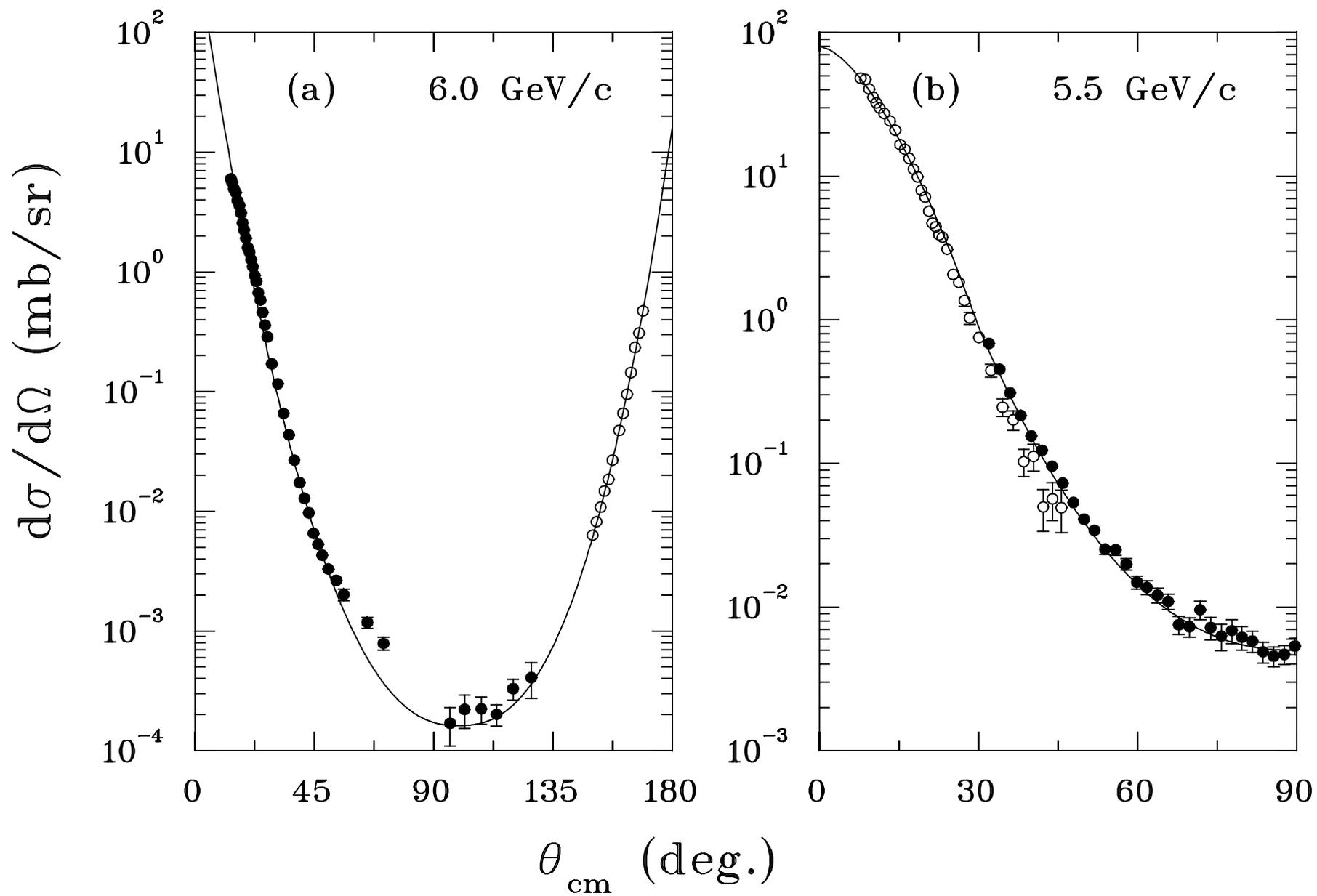
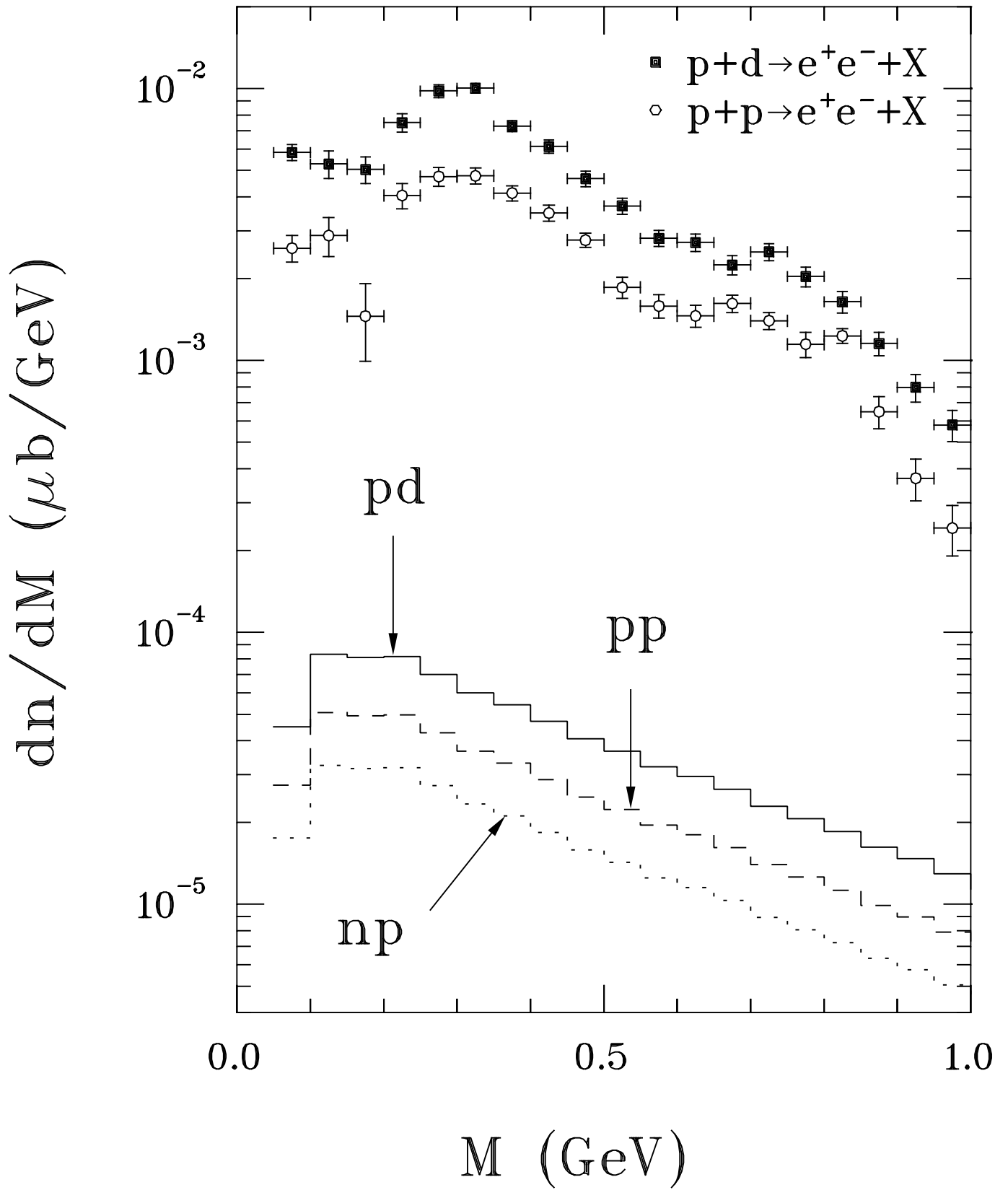
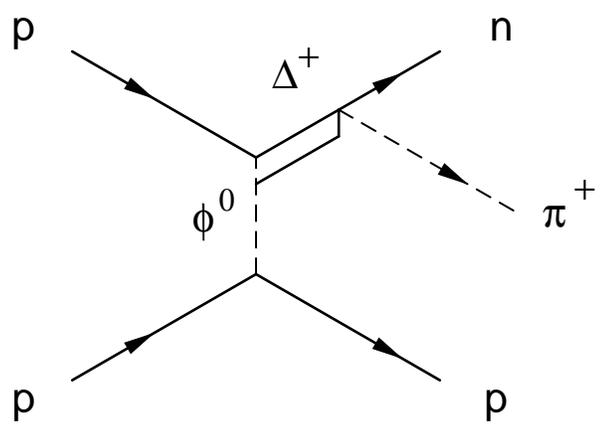
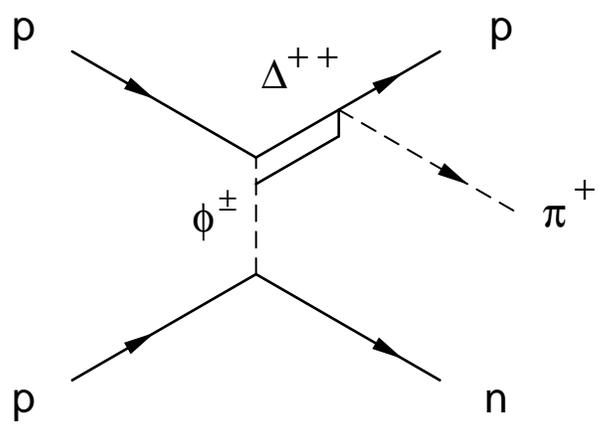
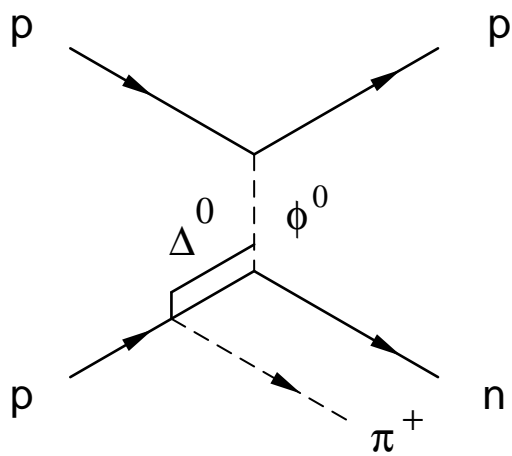
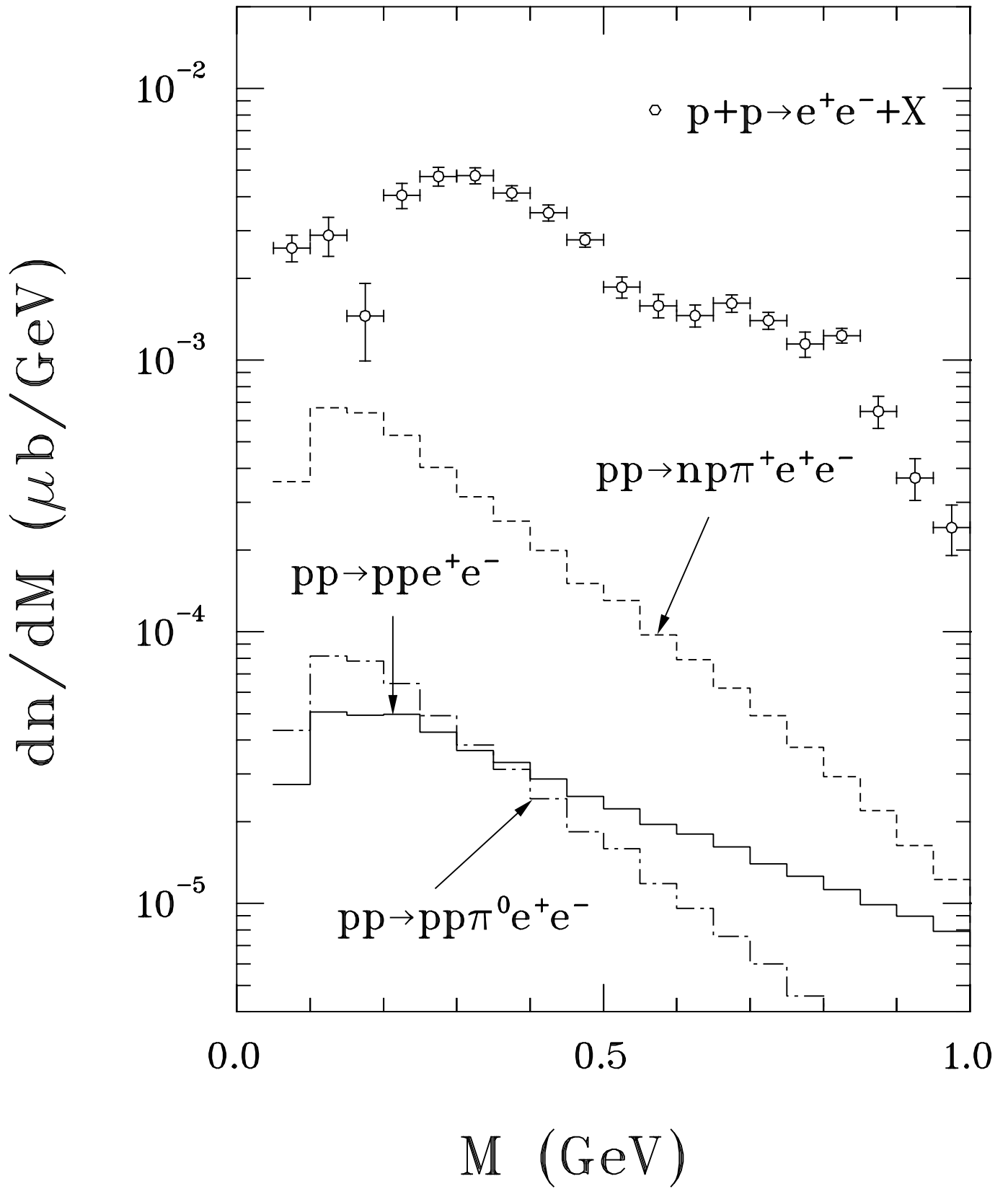
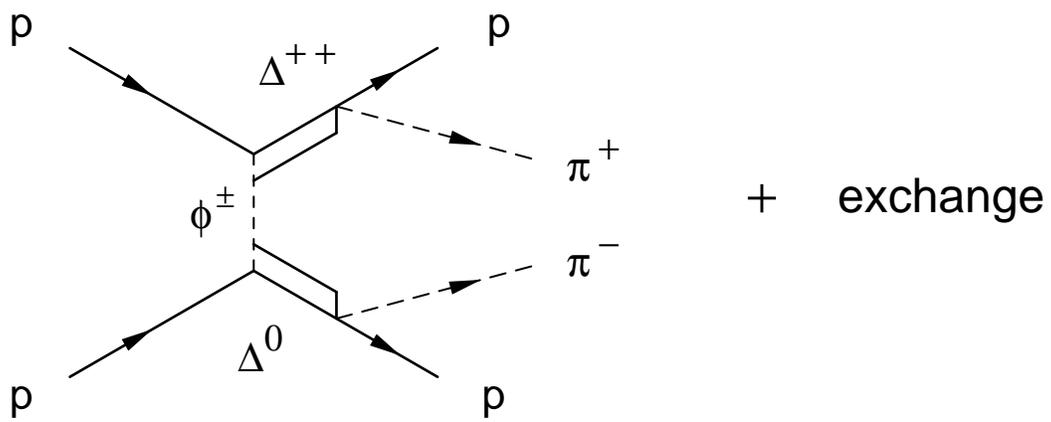
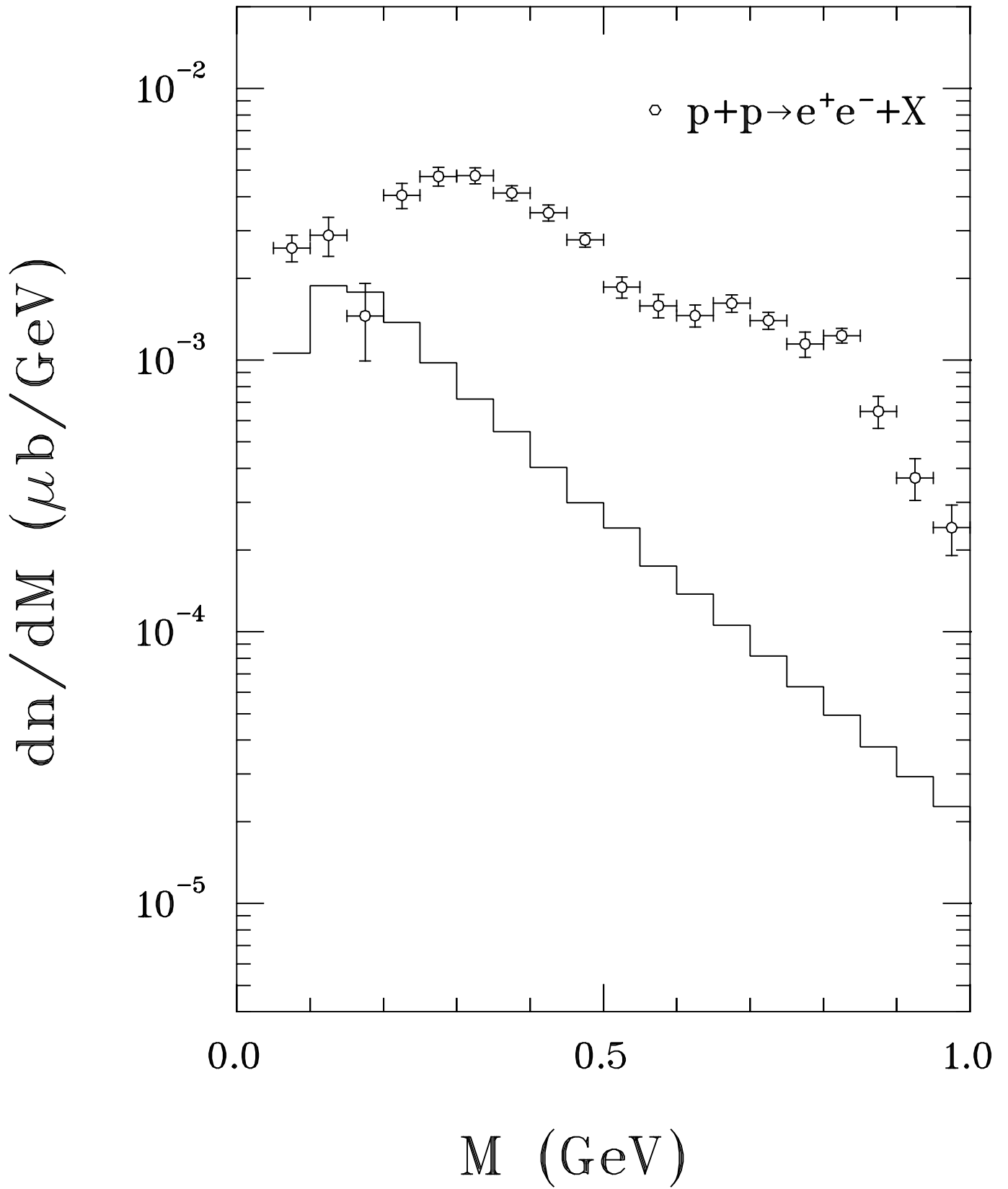
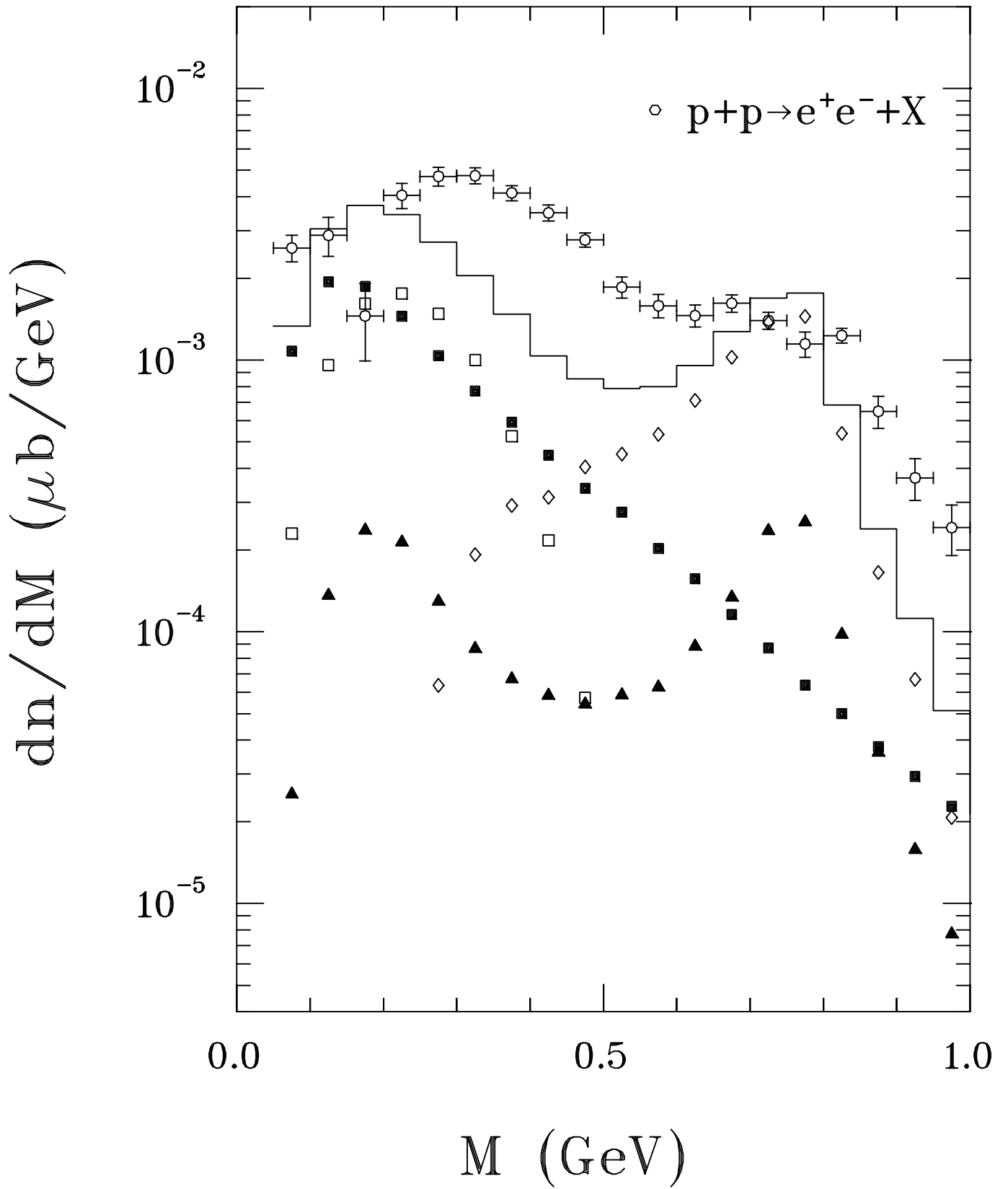
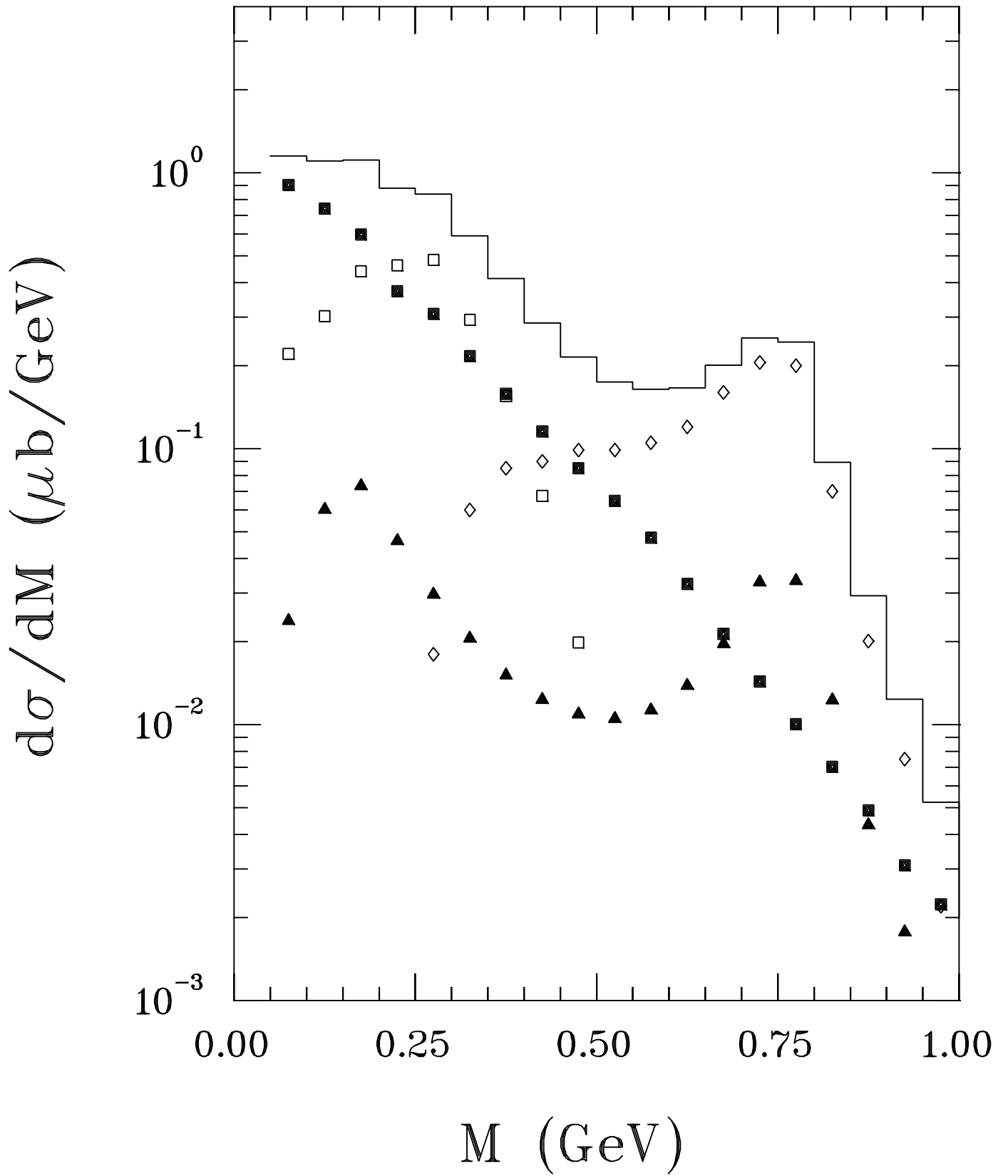
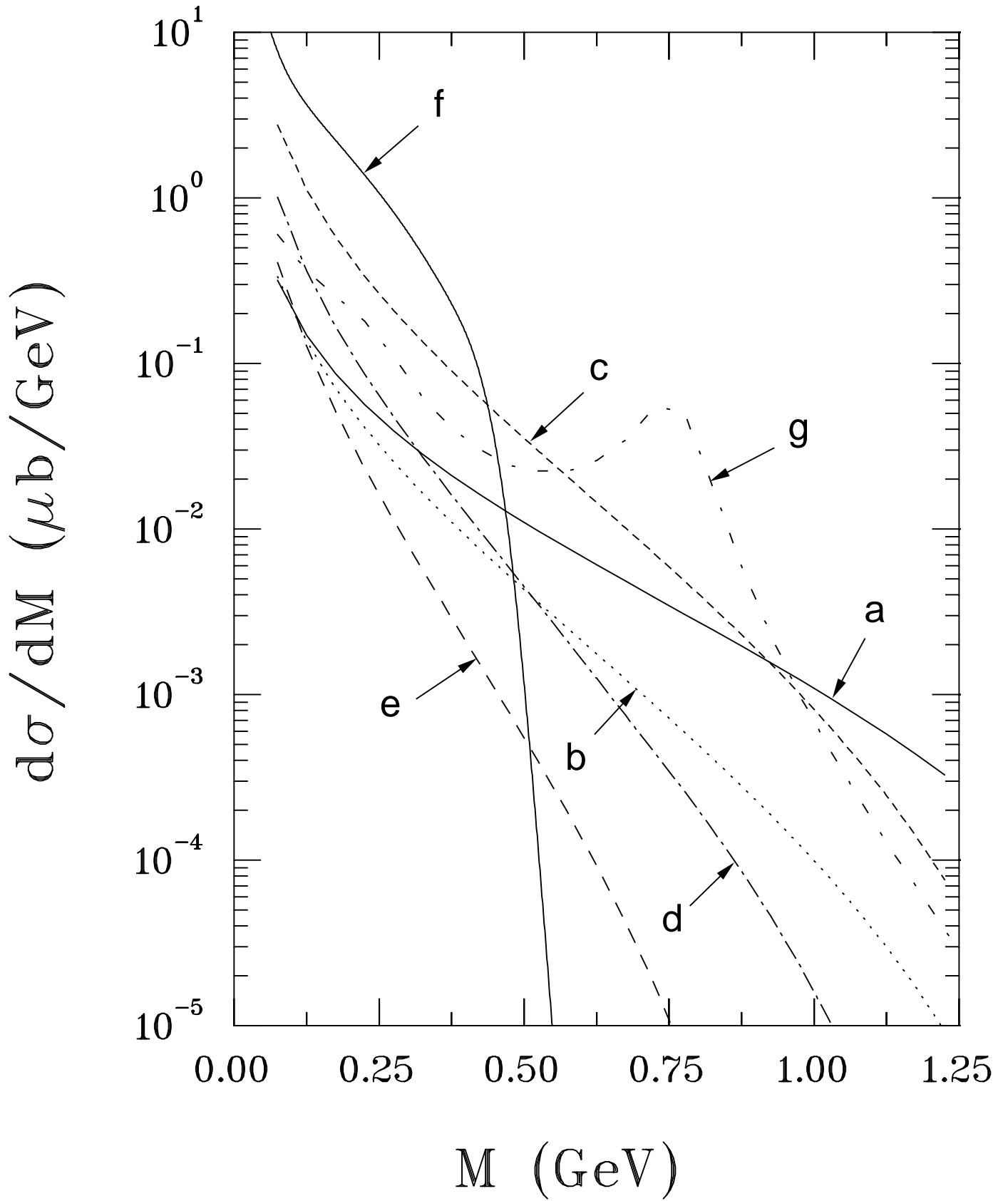
\end{document}